\documentclass[%
 aip,
 amsmath,amssymb,
 reprint,%
]{revtex4-1}
\usepackage[utf8]{inputenc}
\usepackage[T1]{fontenc}
\usepackage{lmodern}
\usepackage{graphicx}
\usepackage{dcolumn}
\usepackage{bm}
\usepackage{mathrsfs}
\usepackage{braket}
\usepackage{siunitx}
\usepackage[colorlinks=true, allcolors=blue]{hyperref}
\usepackage{booktabs}
\usepackage{caption}
\usepackage{subcaption}
\usepackage{multirow}
\usepackage{xcolor}
\usepackage{placeins}
\usepackage{natbib}

\usepackage{doi}

\begin{document}

\preprint{AIP/123-QED}

\title{Projection-Based Memory Kernel Coupling Theory for Quantum Dynamics: A Stable Framework for Non-Markovian Simulations}

\author{Wei Liu}
\affiliation{Department of Chemistry, School of Science and Research Center for Industries of the Future, Westlake University,
Hangzhou, Zhejiang 310030, China}

\author{Rui-Hao Bi}
\affiliation{Department of Chemistry, School of Science and Research Center for Industries of the Future, Westlake University,
Hangzhou, Zhejiang 310030, China}

\author{Yu Su}
\affiliation{Hefei National Research Center for Physical Sciences at the Microscale,
University of Science and Technology of China, Hefei, Anhui 230026, China}

\author{Limin Xu}
\affiliation{Institute for Theoretical Sciences, Westlake University, Hangzhou, 310030, China}

\author{Zhennan Zhou}
\affiliation{Institute for Theoretical Sciences, Westlake University, Hangzhou, 310030, China}

\author{Yao Wang}
\affiliation{Hefei National Research Center for Physical Sciences at the Microscale,
University of Science and Technology of China, Hefei, Anhui 230026, China}

\author{Wenjie Dou}
\email{douwenjie@westlake.edu.cn}
\affiliation{Department of Chemistry, School of Science and Research Center for Industries of the Future, Westlake University,
Hangzhou, Zhejiang 310030, China}
\affiliation{Institute of Natural Sciences, Westlake Institute for Advanced Study, Hangzhou, Zhejiang 310024, China}
\affiliation{Key Laboratory for Quantum Materials of Zhejiang Province, Department of Physics, School of Science and Research
Center for Industries of the Future, Westlake University, Hangzhou, Zhejiang 310030, China}

\date{\today}
\preprint{AIP/123-QED}

\begin{abstract}
We present a projection-based, stability-preserving methodology for computing time correlation functions in open quantum systems governed by generalized quantum master equations with non-Markovian effects. Building upon the memory kernel coupling theory framework, our approach transforms the memory kernel hierarchy into a system of coupled linear differential equations through Mori-Zwanzig projection, followed by spectral projection onto stable eigenmodes to ensure numerical stability. By systematically eliminating unstable modes while preserving the physically relevant dynamics, our method guaranties long-time convergence without introducing artificial damping or \textit{ad hoc} modifications. The theoretical framework maintains mathematical rigor through orthogonal projection operators and spectral decomposition. Benchmark calculations on the spin-boson model show excellent agreement with exact hierarchical equations of motion results while achieving significant computational efficiency. This approach provides a versatile and reliable framework for simulating non-Markovian dynamics in complex systems.
\end{abstract}

\maketitle

\section{Introduction}
The simulation of open quantum systems---quantum subsystems interacting with complex, macroscopic environments---represents a significant challenge in modern theoretical chemistry, condensed matter physics, and quantum information science~\cite{wang2022quantum, tanimura2020numerically}. Such systems govern a diverse variety of phenomena, from ultrafast excitonic energy transfer in photosynthetic complexes~\cite{renger2001ultrafast} to quantum transport and decoherence in nanoscale electronic devices~\cite{chen2024floquet,liu2025enhancement}, where non-Markovian memory effects and system-environment correlations often play essential roles~\cite{li2024toward,han2019stochastic,song2016alternative,segal2008thermal}. At the heart of these processes lies the time correlation function, a fundamental quantity that directly links microscopic quantum dynamics to experimentally accessible spectroscopic observables, transport coefficients, and thermalization rates~\cite{berne1970calculation,bednorz2012nonclassical,cao1996semiclassical,shi2003relationship,wang2022statistical,liu2024polaritons,liu2025absorption,montoya2017approximate,thoss2001generalized}.

Despite significant methodological advances, there is a long-standing dilemma between accuracy and computational tractability in quantum dynamics simulations. On the one hand, numerically exact approaches, such as the hierarchical equations of motion (HEOM)~\cite{tanimura2020numerically}, the dissipaton equation of motion (DEOM)~\cite{yan2016dissipation} and real-time path integral techniques~\cite{makri1995numerical,makarov1994path}, suffer from high computational scaling with system size or memory time, limiting their application to relatively small systems. On the other hand, widely used approximate methods---including perturbative master equations~\cite{redfield1957theory, tokuyama1976statistical,cohen2020green} and mixed quantum-classical schemes~\cite{tully1990molecular, craig2005trajectory, wang2016recent, mannouch2023mapping,wang2025mixed}---often achieve favorable scaling by sacrificing accuracy, especially in non-perturbative regimes, or by violating fundamental physical principles such as detailed balance and positivity.

Projection operator techniques~\cite{nakajima1958quantum, zwanzig1960ensemble, mori1965transport} offer an elegant theoretical framework to tackle this difficulty. By formally partitioning the full dynamics into relevant and irrelevant parts, these methods recast the exact dynamics into a generalized quantum master equation (GQME) for the reduced density matrix~\cite{brian2021generalized}. The memory kernel $K(t)$ within the GQME encodes all non-Markovian effects and typically decays faster than the reduced system's observables, providing a potential avenue for computational leverage~\cite{liu2025enhancement,kidon2018memory}. However, practical computation of $K(t)$ remains formidable, as it requires propagating dynamics in the formally exact orthogonal subspace.

Recent advances in memory kernel coupling theory (MKCT)~\cite{liu2025memory, bi2025universal} have introduced a promising reformulation. MKCT expresses the memory kernel dynamics through a hierarchy of coupled equations of motion, driven by \textit{static} moments. Such a transformation from a dynamical evolution into a static one offers significant conceptual and potential computational advantages. Nevertheless, practical implementations of MKCT are plagued by severe numerical instabilities upon truncation of the hierarchy~\cite{akbari2012challenges,tang2015extended,bai2024hierarchical}. Current stabilization strategies rely on empirically tuned methods, such as Padé approximants~\cite{bi2025universal} or dynamical mode decomposition (DMD)~\cite{liu2025memory,liu2023predicting}, which do not always guarantee convergence.

In this work, we introduce the projection-based memory kernel coupling theory (PMKCT)---a mathematically rigorous stabilization framework that guarantees numerical stability while intrinsically preserving necessary physical constraints for the memory kernel. The core innovation of PMKCT lies in a spectral decomposition of the MKCT generator matrix, segregating its dynamics into stable and unstable subspaces. A subsequent projection step systematically removes the unstable modes responsible for numerical divergence, transforming an \textit{ad hoc} numerical problem into a principled linear algebraic operation with only the stable modes. Our approach completely eliminates empirical parameter tuning and fully retains the computational advantages and formal structure of the original MKCT formalism.

The remainder of this paper is organized as follows. In Section~\ref{sec:theory}, we develop the complete PMKCT framework, detailing the spectral analysis of the MKCT generator and the projection-based stabilization procedure. Section~\ref{sec:results} presents comprehensive numerical benchmarks on the widely studied spin-boson model, comparing PMKCT against established exact methods. Finally, we conclude in Section~\ref{sec:conclusions} by summarizing our key contributions and discussing the broader implications of PMKCT for the future simulation of complex open quantum systems.

\section{Theoretical Framework}
\label{sec:theory}

\subsection{Generalized Quantum Master Equation}
\label{subsec:gqme_foundation}

Our framework builds upon the Mori-Zwanzig projection formalism \cite{mori1965transport}, which provides an exact representation of open quantum system dynamics. For an observable $\hat{A}$, the GQME reads:
\begin{equation}
\dot{C}_{\hat{A}\hat{A}}(t) = \Omega C_{\hat{A}\hat{A}}(t) + \int_0^t d\tau\, K(t-\tau)C_{\hat{A}\hat{A}}(\tau),
\label{eq:gqme_full}
\end{equation}
where $C_{\hat{A}\hat{A}}(t) = \braket{ \hat{A}(t)\hat{A}(0)}$ is the correlation function, $\Omega = \braket{i\mathcal{L}\hat{A} \hat{A}}/\braket{\hat{A}\hat{A}}$ is the frequency parameter, and $K(t) = \braket{i\mathcal{L}e^{it\mathcal{Q}\mathcal{L}}\mathcal{Q}i\mathcal{L}\hat{A} \hat{A}}/\braket{\hat{A}\hat{A}}$ is the memory kernel. The Liouville operator $\mathcal{L}$ acts as $\mathcal{L}\hat{X} = [\hat{H}, \hat{X}]/i\hbar$, and $\mathcal{Q} = I - \mathcal{P}$ is the complement of the projection operator $\mathcal{P}\hat{X} = \braket{\hat{X} \hat{A}}\hat{A}/\braket{\hat{A}\hat{A}}$. The inner product is given by $\braket{\hat{O}_1\hat{O}_2}\equiv\text{Tr}({\hat{O}_1\hat{O}_2}\rho_{\text{ss}})$, where $\rho_{\text{ss}}$ is  the steady state.

The memory kernel $K(t)$ encodes the non-Markovian influence of the environment and typically decays faster than $C_{\hat{A}\hat{A}}(t)$ itself, offering computational advantages. However, computing $K(t)$ remains challenging as it involves the projected propagator $e^{it\mathcal{Q}\mathcal{L}}$ in the orthogonal subspace.

\subsection{Memory Kernel Coupling Theory}
\label{subsec:mkct_framework}

MKCT~\cite{liu2025memory} provides a systematic approach for computing $K(t)$ by introducing auxiliary kernels. Defining the $n$th-order moment:
\begin{equation}
\Omega_n \equiv \braket{(i\mathcal{L})^{n}\hat{A} \hat{A}}/\braket{\hat{A} \hat{A}},
\label{eq:omega_n_def}
\end{equation}
and the $n$th-order auxiliary kernel:
\begin{equation}
K_n(t) \equiv \braket{(i\mathcal{L})^{n}\hat{f}(t)\hat{A}}/\braket{\hat{A}\hat{A}}, \quad \hat{f}(t) = e^{it\mathcal{Q}\mathcal{L}}\mathcal{Q}i\mathcal{L}A,
\label{eq:kernel_n_def}
\end{equation}
the following hierarchy emerges~\cite{liu2025memory}:
\begin{subequations}
\label{eq:mkct_hierarchy}
\begin{align}
\dot{K}_n(t) &= K_{n+1}(t) - \Omega_n K_1(t), \quad n = 1, \dots, N, \label{eq:dynamics} \\
K_n(0) &= \Omega_{n+1} - \Omega_n \Omega_1.\label{eq:initial} 
\end{align}
\end{subequations}

The above equations form a semi-infinite chain that resembles HEOM. In practice, proper truncation has to be introduced. In the following, we set $K_{N+1}(t) =  0$. Note that this formulation elegantly separates static moment computation ($\Omega_n$) from time evolution, with the latter governed by a linear system. In matrix form with $\bm{K}(t) = [K_1(t), \dots, K_N(t)]^\top$:
\begin{equation}
\dot{\bm{K}}(t) = M \bm{K}(t), 
\label{eq:matrix_formulation}
\end{equation}
with the initial condition $\bm{K}(0) = [\Omega_2 - \Omega_1^2, \Omega_3 - \Omega_2\Omega_1, \dots, \Omega_{N+1} - \Omega_N\Omega_1]^\top$, and $M$ has the specific structure
\begin{equation}
M_{ij} = \delta_{i+1,j} - \Omega_i \delta_{j,1},
\label{eq:matrix_elements_full}
\end{equation}
explicitly:
\begin{equation}
M = 
\begin{pmatrix}
-\Omega_1 & 1 & 0 & \cdots & 0 \\
-\Omega_2 & 0 & 1 & \cdots & 0 \\
\vdots & \vdots & \ddots & \ddots & \vdots \\
-\Omega_{N-1} & 0 & \cdots & 0 & 1 \\
-\Omega_N & 0 & \cdots & 0 & 0
\end{pmatrix}_{N\times N}.
\label{eq:matrix_structure}
\end{equation}

The formal solution is $\bm{K}(t) = \exp(Mt)\bm{K}(0)$, with $K_1(t)$ corresponding to the physical memory kernel in Eq.~\eqref{eq:gqme_full}.

\subsection{Stability Challenge and Projection-Based Solution}
\label{subsec:stabilization}

While Eq.~\eqref{eq:matrix_formulation} provides an exact representation for infinite $N$, truncation to finite $N$ introduces eigenvalues with positive real parts, causing exponential divergence. This numerical instability represents a fundamental limitation of practical MKCT implementations. In previous study, we have used DMD~\cite{liu2025memory} or Padé~\cite{bi2025universal} to remove the divergence. In the following, we introduce the projection based method.  

Our projection-based approach addresses this by systematically decomposing the dynamics into stable and unstable components. Consider the spectral decomposition $M = V\Lambda V^{-1}$ with eigenvalues $\lambda_i$. We classify:
\begin{align}
\mathcal{S} &= \{\lambda_i : \Re(\lambda_i) < 0\} \quad &\text{(stable)}, \\
\mathcal{N} &= \{\lambda_i : \Re(\lambda_i) = 0\} \quad &\text{(neutral)}, \\
\mathcal{U} &= \{\lambda_i : \Re(\lambda_i) > 0\} \quad &\text{(unstable)}.
\end{align}

Let $V_S$ contain eigenvectors for $\mathcal{S} \cup \mathcal{N}$. The orthogonal projection onto $\mathcal{V}_S = \text{span}(V_S)$ is~\cite{barata2012moore,macausland2014moore}:
\begin{equation}
P_S = V_S (V_S^\dagger V_S)^{-1} V_S^\dagger.
\label{eq:projection}
\end{equation}

Applying this projection yields the stabilized system:
\begin{subequations}
\begin{align}
\dot{\bm{K}}(t) &= M_S \bm{K}(t), \\
M_S &= P_S M P_S, 
\end{align}
\end{subequations}
with solution $\bm{K}(t) = \exp(M_S t)\bm{K}(0)$.

For the projected system, $\sigma(M_S) \subseteq \{\lambda \in \mathbb{C} : \Re(\lambda) \leq 0\}$, and any eigenvalues with $\Re(\lambda) = 0$ are semisimple. $M_S$ acts invariantly on $\mathcal{V}_S$ with spectrum $\sigma(M_S) = \sigma(M) \cap (\mathcal{S} \cup \mathcal{N})$. The orthogonal projection ensures neutral eigenvalues are non-defective.

The reason why one can project out the unstable modes is tied to the breakdown of time-reversal  symmetry. For open quantum systems, the time reversibility is violated. However, such time reversibility is preserved in the original MKCT. By removing the unstable modes, one introduces breakdown of time-reversal  symmetry in MKCT. Notice that introducing breakdown of time-reversal  symmetry is also essential in DEOM. In DEOM, the Wick theorem is used differently for positive time and negative time. \cite{yan2016dissipation} Below, we show that this projected based method works in practice. 

\subsection{Numerical Implementation Considerations}
\label{ssec:numerical_considerations}

For improved numerical conditioning with large $N$, particularly when $\Omega_n$ span multiple orders of magnitude, we employ variable rescaling~\cite{trefethen2022numerical, golub2013matrix,butcher2016numerical}. Two effective schemes are presented below, each with its distinct matrix representation.

\subsubsection{Factorial Scaling Scheme}
\label{subsubsec:factorial_scaling}

Define scaled variables $\tilde{K}_n(t) = K_n(t)/(n!\Lambda^{n-1})$ with characteristic frequency $\Lambda$. The transformed hierarchy becomes:
\begin{equation}
\dot{\tilde{K}}_n(t) = (n+1)\Lambda \tilde{K}_{n+1}(t) - \frac{\Omega_n}{n!\Lambda^{n-1}}\tilde{K}_1(t),
\label{eq:factorial_scaling}
\end{equation}
with $\tilde{K}_{N+1}(t) = 0$.
In matrix form with $\tilde{\bm{K}}(t) = [\tilde{K}_1(t), \dots, \tilde{K}_N(t)]^\top$:
\begin{equation}
\frac{d}{dt}\tilde{\bm{K}}(t) = \tilde{M}^{(1)} \tilde{\bm{K}}(t),
\label{eq:matrix_factorial}
\end{equation}
where the $N \times N$ matrix $\tilde{M}^{(1)}$ has elements:
\begin{equation}
\tilde{M}^{(1)}_{ij} = (i+1)\Lambda\delta_{i+1,j} - \frac{\Omega_i}{i!\Lambda^{i-1}}\delta_{j,1}.
\label{eq:matrix1_elements}
\end{equation}

\subsubsection{Power-Law Scaling Scheme}
\label{subsubsec:powerlaw_scaling}

For $\tilde{K}_n(t) = K_n(t)/\Lambda^{n-1}$, we obtain:
\begin{equation}
\dot{\tilde{K}}_n(t) = \Lambda \tilde{K}_{n+1}(t) - \frac{\Omega_n}{\Lambda^{n-1}}\tilde{K}_1(t),
\label{eq:powerlaw_scaling}
\end{equation}
with $\tilde{K}_{N+1}(t) = 0$.
The corresponding matrix $\tilde{M}^{(2)}$ has elements:
\begin{equation}
\tilde{M}^{(2)}_{ij} = \Lambda\delta_{i+1,j} - \frac{\Omega_i}{\Lambda^{i-1}}\delta_{j,1}.
\label{eq:matrix2_elements}
\end{equation}

\subsubsection{Projection Framework Compatibility}

The projection stabilization procedure established in Section~\ref{subsec:stabilization} applies directly to both rescaled formulations. For $\tilde{M}^{(k)} \in \{\tilde{M}^{(1)}, \tilde{M}^{(2)}\}$, we compute stabilized matrices via:

\begin{equation}
\tilde{M}^{(k)}_S = P_S^{(k)} \tilde{M}^{(k)} P_S^{(k)},
\end{equation}
where $P_S^{(k)}$ denotes the projection operator constructed from stable eigenvectors of $\tilde{M}^{(k)}$. This maintains mathematical consistency across all formulations while improving numerical conditioning.

\subsubsection{Physical Kernel Extraction}

A key advantage of both rescaling schemes is the trivial recovery of the physical memory kernel:
\begin{equation}
K_1(t) = \tilde{K}_1(t),
\label{eq:kernel_recovery}
\end{equation}
since scaling factors cancel precisely for $n=1$. No inverse transformation is required, providing direct access to the physically meaningful quantity. Higher-order auxiliary kernels ($n>1$) serve exclusively as numerical intermediaries to ensure stability; they possess no independent physical significance and need not be recovered explicitly.





\section{Results and Discussion}
\label{sec:results}

We validate our PMKCT framework on the spin-boson model with Ohmic spectral density, described by the Hamiltonian $H_{\text{}}=H_{\text{S}}+H_{\text{SB}}+h_{\text{B}}=H_{\text{S}}+ \hat{Q}\hat{F} + h_{\text{B}}$,
where $H_{\text{S}}$ and $h_{\text{B}}$ are the system Hamiltonian and the bath Hamiltonian, and
$H_{\text{SB}}$ is the coupling between the system and bath. $\hat{Q}$ is a system operator $\sigma_x$ and $\hat{F}$ is an environment operator, which are both Hermitian. We have 
\begin{equation}
\begin{aligned}
        H_{\text{S}} &= \frac{1}{2}\Delta\sigma_z + \epsilon\sigma_x,\\
        h_{\text{B}} &= \sum_{j} \frac{1}{2} \omega_j (\hat{p}^2_j+\hat{x}^2_j),\\
        \hat{F} &= \sum_{j}c_j\hat{x}_j.
\end{aligned}  
\end{equation}
Here, $\Delta$ is the unit energy, which corresponds to the energy difference between the two sites. $\epsilon$ represents the tunneling matrix element, which is assumed to be static. And $\sigma_i$ corresponds to the $i^{\text{th}}$ Pauli matrix. $p_j$, $x_j$, and $\omega_j$ are the momenta, coordinates, and frequency for the $j^{\text{th}}$ harmonic oscillator, respectively. $c_j$ is the coupling constant that describes the strength of the interaction between the system and the $j^{\text{th}}$ oscillator. The initial state $\hat{\sigma}_0 = |0\rangle\langle 0|$. We compute the dipole autocorrelation function $C_{\hat{\mu}\hat{\mu}}(t) = \langle \hat{\sigma}_x(t)\hat{\sigma}_x(0)\rangle$, where the absorption lineshape is obtained via Fourier transform:
\begin{equation}
I(\omega) \propto \Re \int_0^\infty dt\, C_{\hat{\mu}\hat{\mu}}(t)e^{i\omega t}.
\label{eq:lineshape}
\end{equation}

The system-bath interaction is fully characterized by the Ohmic spectral density,
\begin{equation}
J(\omega) = 2\gamma \omega e^{-|\omega|/\omega_D},
\label{eq:ohmic}
\end{equation}
where $\gamma$ controls coupling strength and $\omega_D$ is the cutoff frequency. Throughout, we set $\hbar = 1$.

\subsection{Stability Analysis and Eigenvalue Spectra}
\label{subsec:eigenvalue_analysis}

\begin{figure*}[htbp]  
\centering
\includegraphics[width=\textwidth]{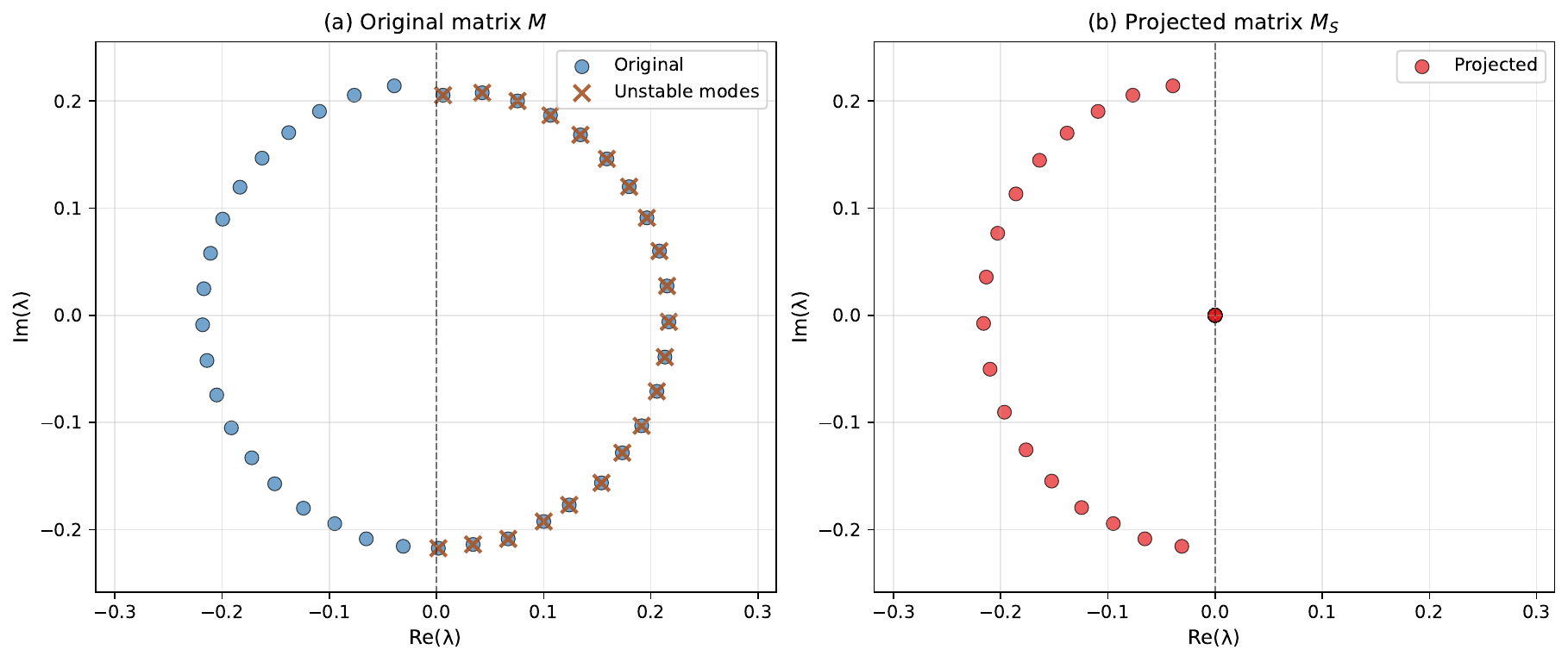} 
\caption{Eigenvalue spectra for the spin-boson model with $\Delta = 20$, $\gamma = 0.5$, $\omega_D = 1$, $\beta = 5$, $N=40$. Power-law scaling is used with $\Lambda=100$. (a) Original matrix $M$ shows 21 eigenvalues with $\Re(\lambda) > 0$ (crosses). (b) After projection, all eigenvalues satisfy $\Re(\lambda) \leq 0$.}
\label{fig:eigenvalue_comparison}
\end{figure*}

In Fig.~\ref{fig:eigenvalue_comparison}, we demonstrate the stabilization effect of PMKCT for the spin-boson model. The original matrix $M$ with $N=40$ exhibits 21 eigenvalues with positive real parts (Fig.~\ref{fig:eigenvalue_comparison}a), which would cause exponential divergence in direct integration. After applying our projection procedure, all eigenvalues of $M_S$ reside in the left half-plane (Fig.~\ref{fig:eigenvalue_comparison}b), guaranteeing asymptotic stability.

The number of unstable modes increases with truncation order $N$, as shown in Table~\ref{tab:unstable_modes}. This highlights the importance of stabilization for accurate long-time simulations.

\begin{table}[htbp]
\centering
\caption{Unstable modes in original $M$ versus truncation order $N$}
\label{tab:unstable_modes}
\begin{tabular}{lccccc}
\toprule
$N$ & 10 & 20 & 30 & 40  \\
\midrule
Unstable modes & 5 & 10 & 15 & 21  \\
$\max_{\lambda \in \mathcal{U}} \Re(\lambda)$ & 0.199 & 0.202 & 0.206 & 0.217\\
$\min_{\lambda \in \mathcal{U}} \Re(\lambda)$ & 0.057 & 0.031 & 0.026 & 0.006\\
\bottomrule
\end{tabular}
\end{table}

\subsection{Memory Kernel and Correlation Function Dynamics}
\label{subsec:kernel_dynamics}

\begin{figure*}[htbp]
\centering
\includegraphics[width=\textwidth]{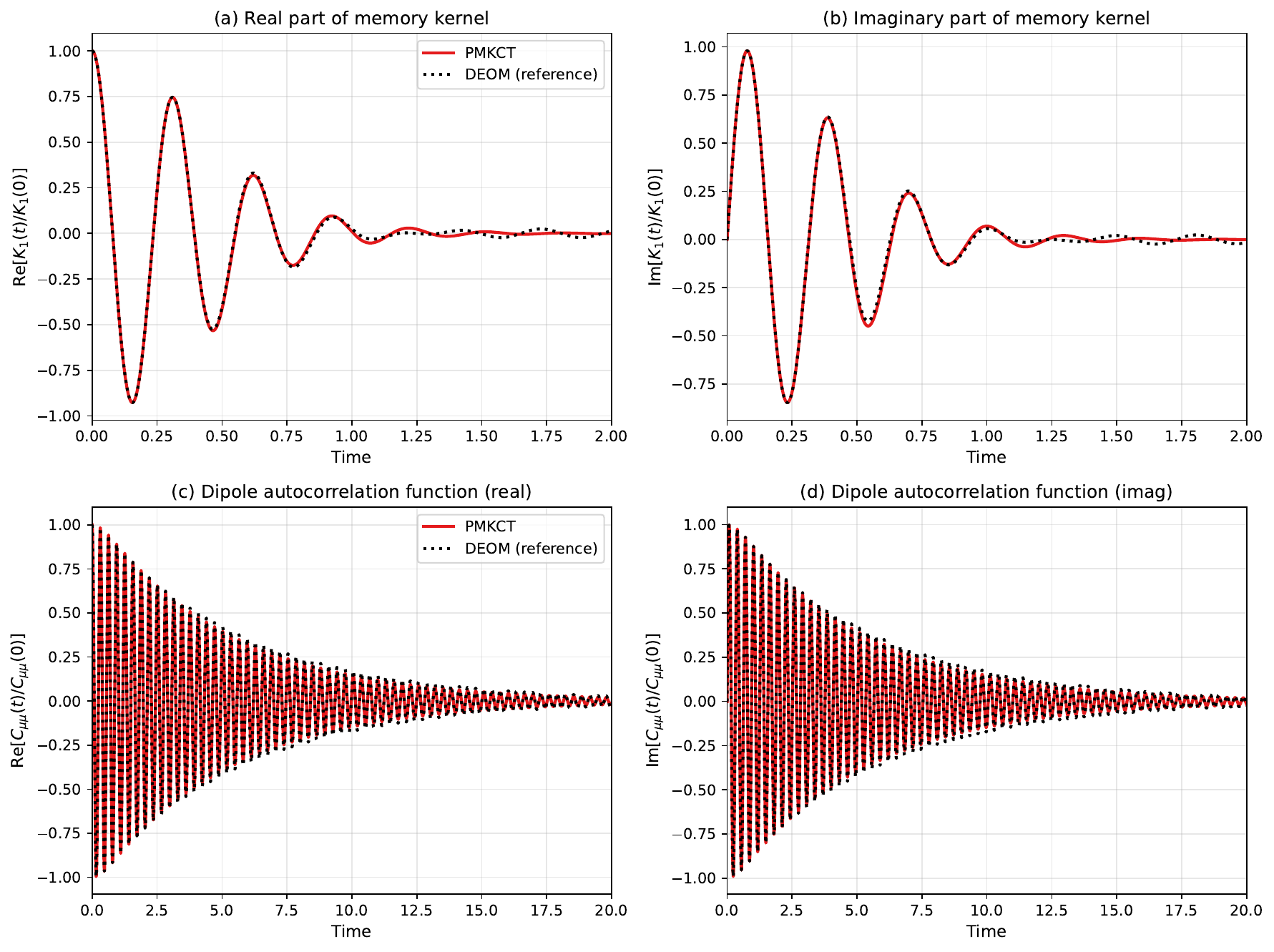}
\caption{Time evolution for the spin-boson model (parameters as in Fig.~\ref{fig:eigenvalue_comparison}). Solid lines: PMKCT with $N=40$. Dotted lines: Reference DEOM calculation.}
\label{fig:dynamics_comparison}
\end{figure*}

In Fig.~\ref{fig:dynamics_comparison}, we compare PMKCT results with numerically exact dissipaton equation of motion (DEOM)~\cite{yan2016dissipation} calculations. For the memory kernel $K_1(t)$ (Fig.~\ref{fig:dynamics_comparison}a and Fig.~\ref{fig:dynamics_comparison}b), PMKCT achieves excellent agreement with DEOM across the entire time range. 

The correlation function $C_{\hat{\mu}\hat{\mu}}(t)$ (Fig.~\ref{fig:dynamics_comparison}c and Fig.~\ref{fig:dynamics_comparison}d) exhibits rapid oscillations due to the large energy gap ($\Delta = 20$). PMKCT captures these oscillations accurately.

A key observation is that $K_1(t)$ decays significantly faster than $C_{\hat{\mu}\hat{\mu}}(t)$, decaying to near-zero by $t \approx 2$, while $C_{\hat{\mu}\hat{\mu}}(t)$ maintains substantial amplitude beyond $t = 10$. This rapid kernel decay is exploited in both PMKCT and MKCT frameworks for efficient long-time simulations.

\subsection{Convergence with Truncation Order}
\label{subsec:convergence}

\begin{figure*}[htbp]
\centering
\begin{subfigure}{0.48\textwidth}
    \includegraphics[width=\linewidth]{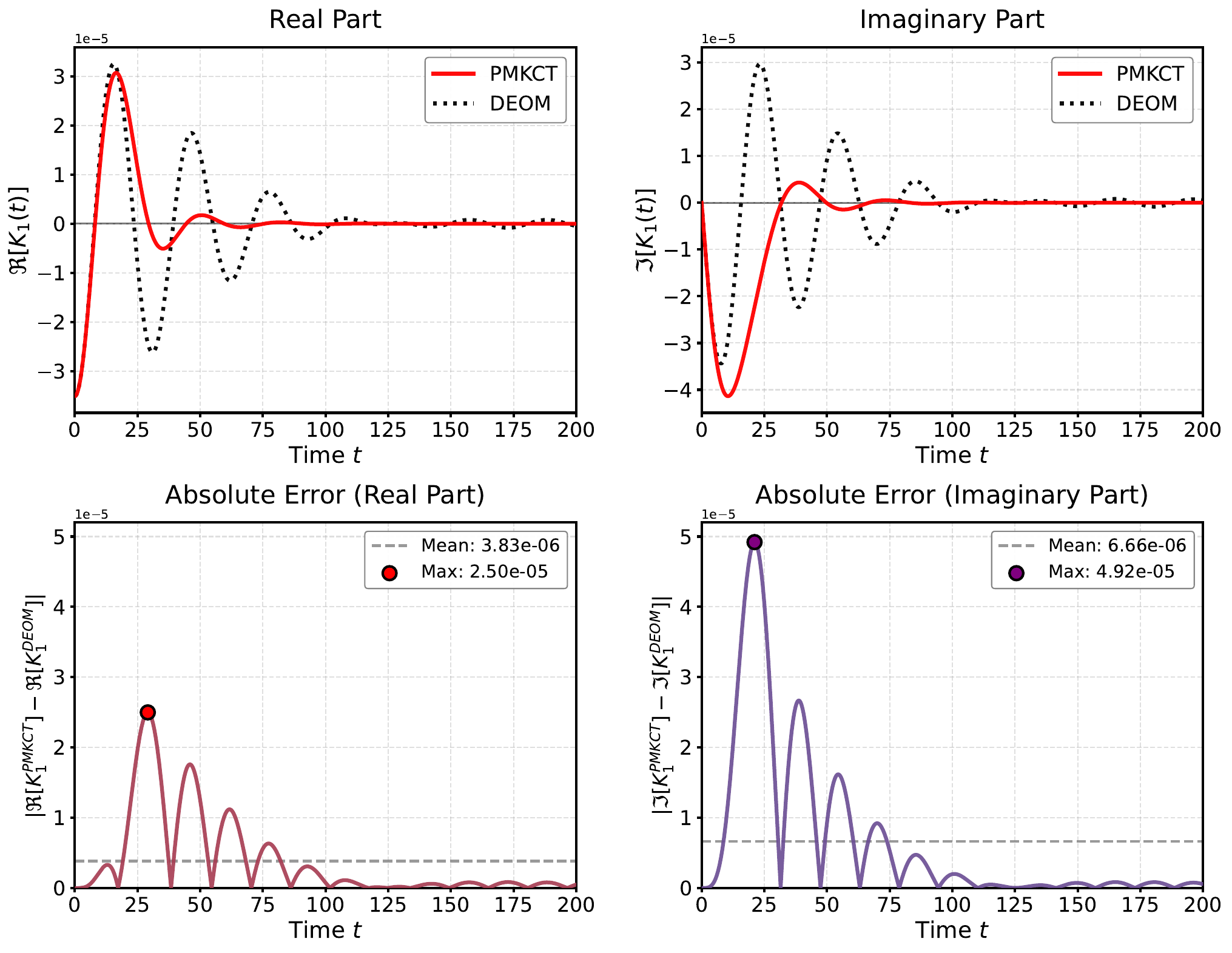}
    \caption{$N=10$}
    \label{fig:conv_N10}
\end{subfigure}
\hfill
\begin{subfigure}{0.48\textwidth}
    \includegraphics[width=\linewidth]{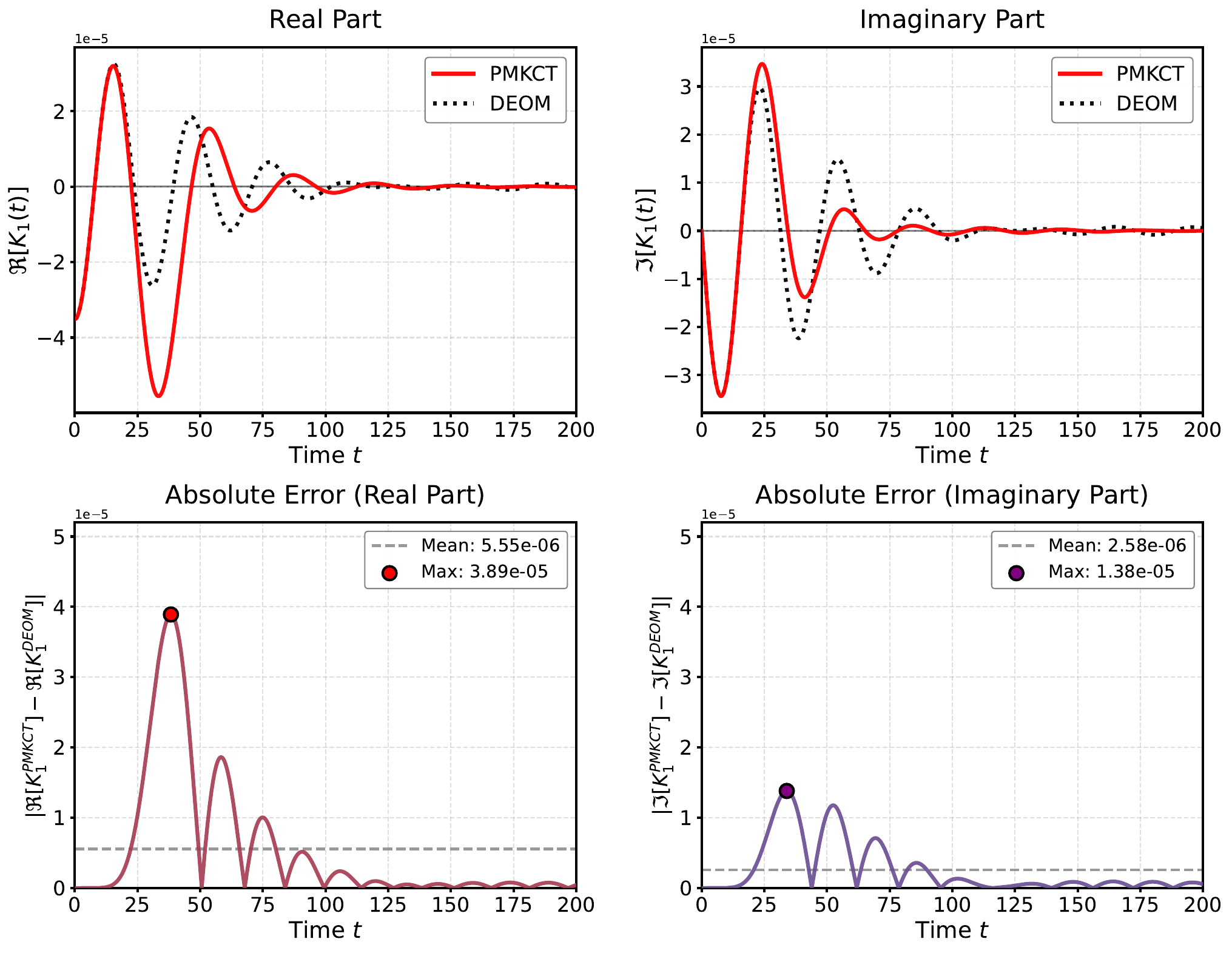}
    \caption{$N=20$}
    \label{fig:conv_N20}
\end{subfigure}
\vskip\baselineskip
\begin{subfigure}{0.48\textwidth}
    \includegraphics[width=\linewidth]{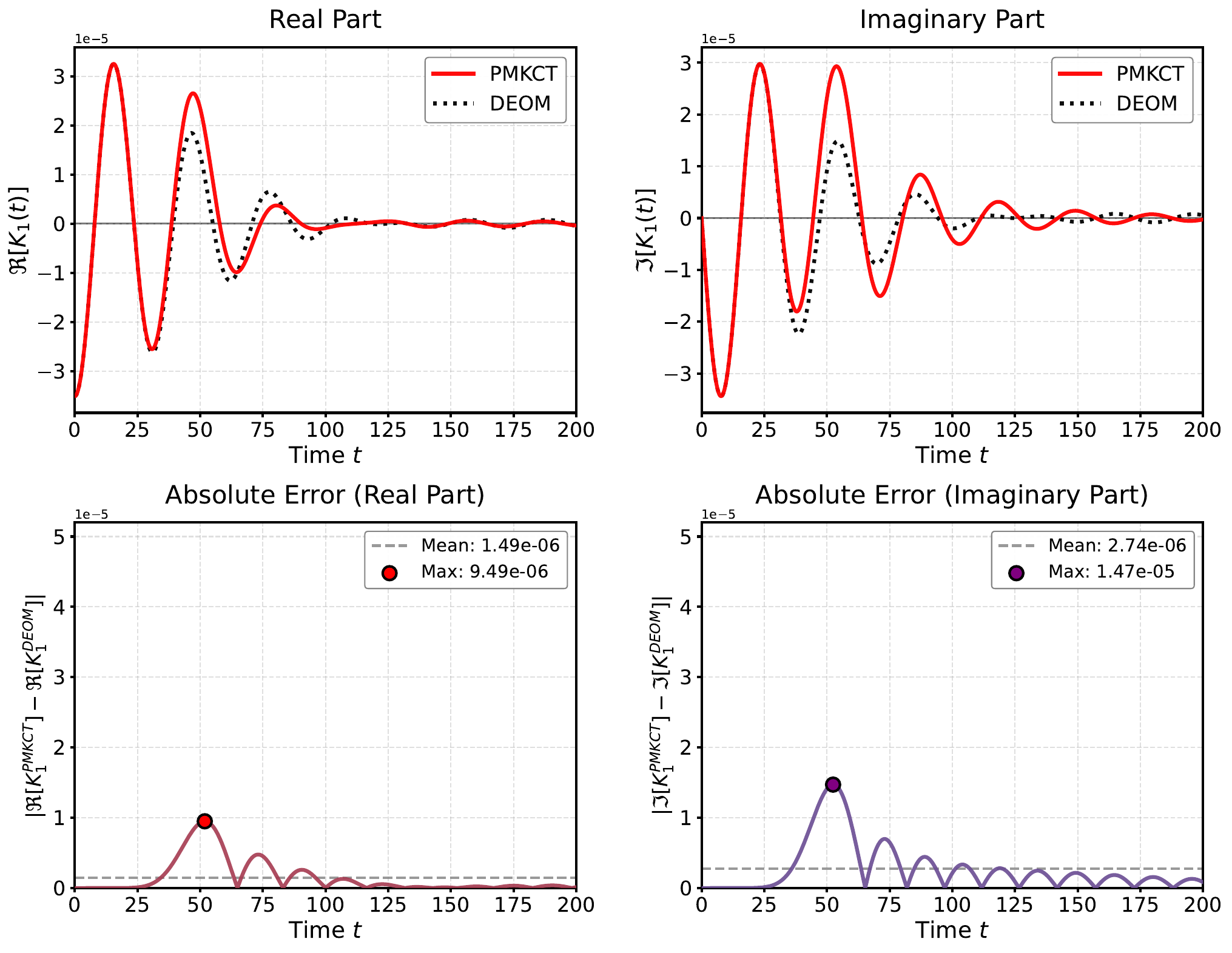}
    \caption{$N=30$}
    \label{fig:conv_N30}
\end{subfigure}
\hfill
\begin{subfigure}{0.48\textwidth}
    \includegraphics[width=\linewidth]{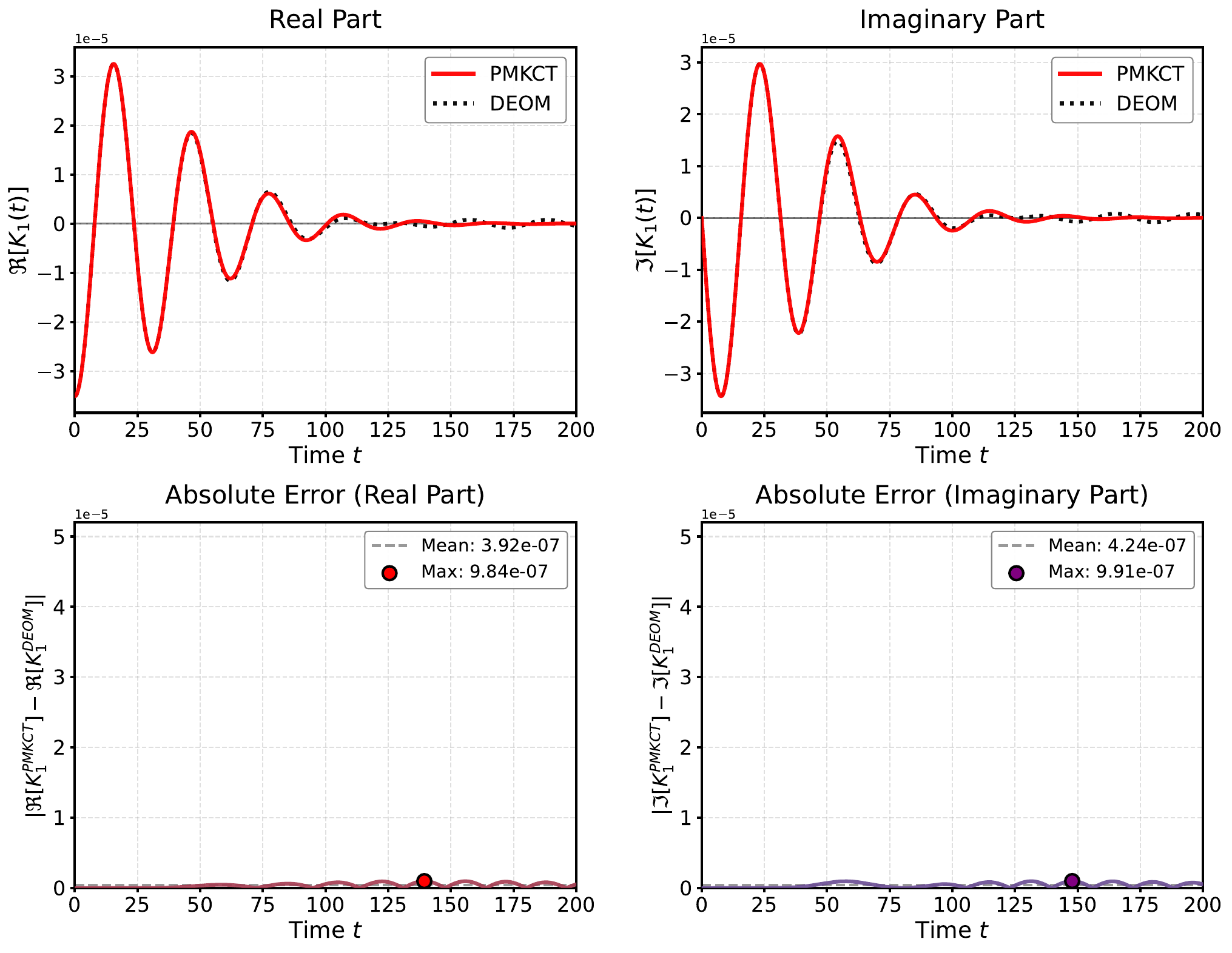}
    \caption{$N=40$}
    \label{fig:conv_N40}
\end{subfigure}
\caption{Convergence of $K_1(t=5)$ with truncation order $N$ in PMKCT for different values: (a) $N=10$, (b) $N=20$, (c) $N=30$, (d) $N=40$. }
\label{fig:convergence}
\end{figure*}

In Fig.~\ref{fig:convergence}, we investigate the convergence of the PMKCT method as the truncation order $N$ increases. The error of memory kernel decreases with larger $N$, which aligns with the theoretical expectations for continued fraction truncation. For this parameter set, an absolute error within $10^{-7}$ is achieved at $N = 40$.

\subsection{Frequency-Domain Analysis}
\label{subsec:frequency_domain}

\begin{figure*}[htbp]
\centering
\includegraphics[width=\textwidth]{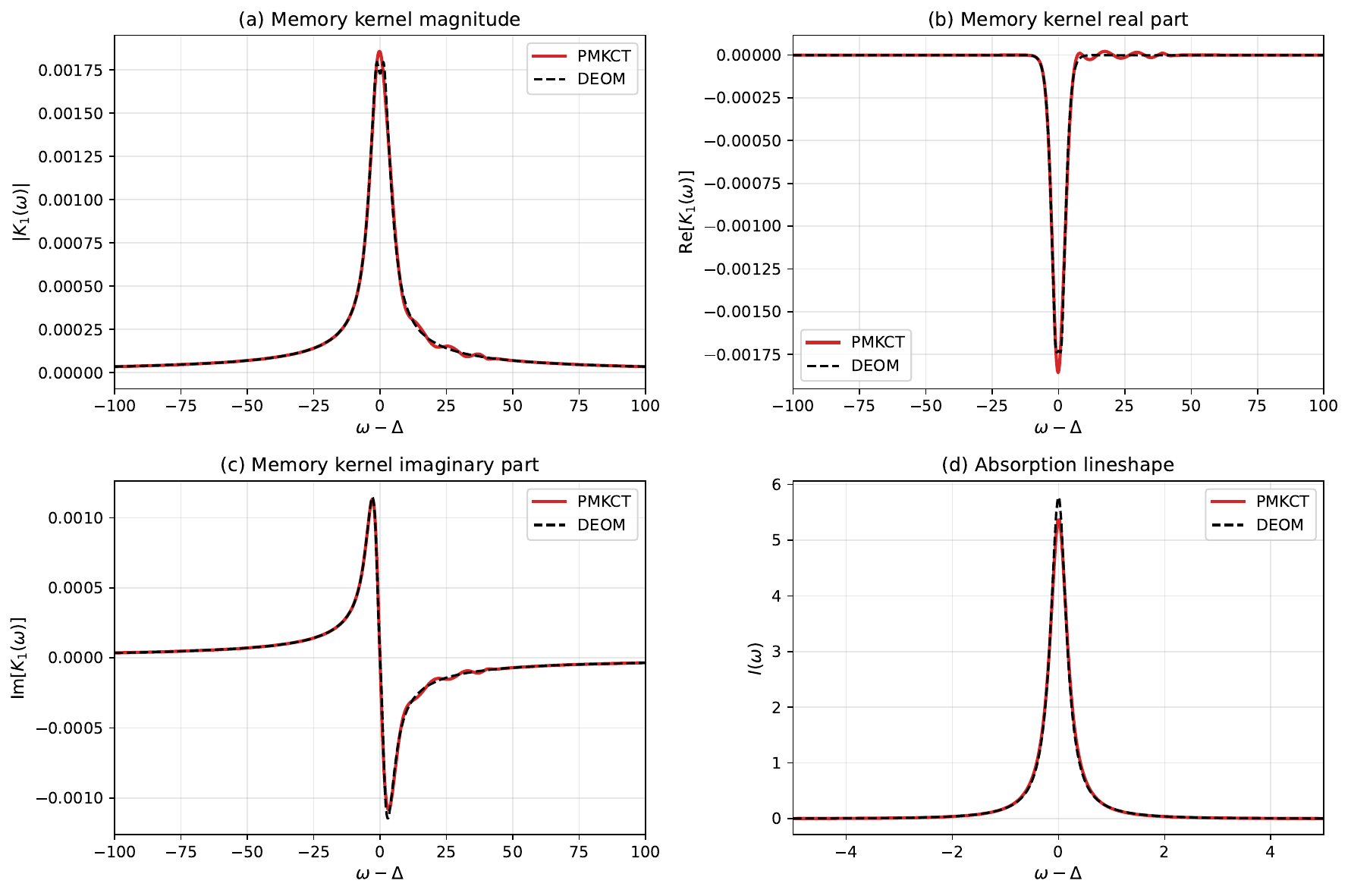}
\caption{Frequency-domain results. (a) Memory kernel magnitude. (b) Memory kernel real part. (c) Memory kernel imaginary part. (d) Absorption lineshape. Parameters as in previous figures.}
\label{fig:frequency_domain}
\end{figure*}

In Fig.~\ref{fig:frequency_domain}, we present frequency-domain results. The memory kernel spectrum $K_1(\omega)$ (Figs.~\ref{fig:frequency_domain}a, ~\ref{fig:frequency_domain}b and ~\ref{fig:frequency_domain}c) displays a broad distribution centered around $\omega \approx \Delta = 20$, with significant weight extending to higher frequencies. This broad spectrum reflects the non-Markovian character of the dynamics. Due to the projection operation, the memory kernel of PMKCT exhibits oscillations in the frequency range of 10–40 (on the x-axis). However, these oscillations do not affect the peak positions in the resulting absorption lineshape.

The absorption lineshape $I(\omega)$ (Fig.~\ref{fig:frequency_domain}d) shows the expected peak structure, with PMKCT accurately reproducing the DEOM reference. The slight asymmetry arises from the finite temperature ($\beta = 5$) and Ohmic bath characteristics.

\section{Conclusions}
\label{sec:conclusions}

We have developed a projection-based memory kernel coupling theory (PMKCT) that provides a stable, efficient framework for simulating non-Markovian quantum dynamics. Building upon the MKCT formalism, our approach employs spectral projection techniques to guarantee numerical stability by construction while maintaining physical consistency. The principal contributions are: establishing a rigorous mathematical framework based on orthogonal projection operators and spectral theory and ensuring guaranteed asymptotic stability through systematic elimination of unstable modes.

PMKCT enables accurate long-time simulations of complex open quantum systems, as demonstrated on the spin-boson model with linear couplings. The method provides a systematic alternative to empirical truncation schemes like Padé approximants, offering guaranteed convergence without parameter tuning. Future extensions will target larger systems and more complex interactions, with applications spanning quantum materials, chemical dynamics, and biological systems.

\section*{Acknowledgments}

We thank Jian-Guo Liu and Xu'an Dou for useful discussions. This work was supported by the National Natural Science Foundation of China (Grant Nos. 22361142829, 22273075) and the Zhejiang Provincial Natural Science Foundation (Grant No. XHD24B0301). Computational resources were provided by the Westlake University Supercomputer Center.

\section*{Author Declarations}

\subsection*{Conflict of Interest}
The authors have no conflicts to disclose.

\bibliography{ref}

\end{document}